\documentclass{IEEEoj-data}
\usepackage{cite}
\usepackage{amsmath,amssymb,amsfonts}
\usepackage{algorithmic}
\usepackage{graphicx,color}
\usepackage{subcaption} 
\usepackage{textcomp}
\usepackage{hyperref}
\usepackage{balance}
\usepackage{soul}
\def\BibTeX{{\rm B\kern-.05em{\sc i\kern-.025em b}\kern-.08em
    T\kern-.1667em\lower.7ex\hbox{E}\kern-.125emX}}
\AtBeginDocument{\definecolor{ojcolor}{cmyk}{0.93,0.59,0.15,0.02}}
\usepackage{verbatim}
\usepackage{xcolor}
\usepackage{booktabs}
\usepackage{multirow}
\usepackage{makecell}

\begin{document}

\title{\textcolor{black}{Collection:}  \textcolor{ieeedata}{\textit{UAV-Based Wireless Multi-modal Measurements from AERPAW Autonomous Data Mule (AADM) Challenge in Digital Twin and Real-World Environments}}}

\author{Md Sharif Hossen\authorrefmark{1},
Cole Dickerson\authorrefmark{1}, Ozgur Ozdemir\authorrefmark{1}, Anil Gurses\authorrefmark{1}, Mohamed Rabeek Sarbudeen\authorrefmark{1}, Thomas Zajkowski\authorrefmark{1}, 
Ahmed Manavi Alam\authorrefmark{1}, 
Everett Tucker\authorrefmark{1}, 
William Bjorndahl\authorrefmark{2}, Fred Solis\authorrefmark{2}, 
Sadaf Javed\authorrefmark{3}, 
Anirudh Kamath\authorrefmark{4}, Xiangyao Tang\authorrefmark{4}, 
Joarder Jafor Sadique\authorrefmark{5}, 
Kevin Liu Hermstein\authorrefmark{6}, 
Kaies Al Mahmud\authorrefmark{7},
Jose Angel Sanchez Viloria\authorrefmark{8}, 
Skyler Hawkins\authorrefmark{9},
Yuqing Cui\authorrefmark{10}, Annoy Dey\authorrefmark{10}, 
Yuchen Liu\authorrefmark{1}, Ali Gurbuz\authorrefmark{1}, Joseph Camp\authorrefmark{2}, Rizwan Ahmad\authorrefmark{3}, Jacobus van der Merwe\authorrefmark{4}, Ahmed Ibrahim Mohamed\authorrefmark{5}, Gil Zussman\authorrefmark{6}, Mehmet Kurum\authorrefmark{7}, Namuduri Kamesh\authorrefmark{9}, Zhangyu Guan\authorrefmark{10}, Dimitris Pados\authorrefmark{8}, George Sklivanitis\authorrefmark{8}, Ismail Guvenc\authorrefmark{1}, Mihail Sichitiu\authorrefmark{1}, Magreth Mushi\authorrefmark{1}, Rudra Dutta\authorrefmark{1} 
}


\affil{\scriptsize
North Carolina State University, Raleigh, NC, USA; 
$^{2}$Southern Methodist University, TX, USA;
$^{3}$National University of Sciences and Technology, Pakistan; 
$^{4}$University of Utah, UT, USA; 
$^{5}$Florida International University, FL, USA;
$^{6}$Columbia University, NY, USA; 
$^{7}$The University of Georgia, GA, USA; 
$^{8}$Florida Atlantic University, FL, USA; 
$^{9}$The University of North Texas, TX, USA; \\
$^{10}$University of Minnesota, MN, USA
}

\corresp{CORRESPONDING AUTHOR: Md Sharif Hossen (e-mail: mhossen@ncsu.edu).}
\authornote{This work is supported in part by the  NSF under Grant CNS-1939334, CNS-2450593, 2332661, OAC-2512931, 2346555, CNS-2144297, CNS-2148128, CNS-2332662, CNS-2117822, EEC-2133516, CNS-2440756, SWIFT-2229563, DGE-2137100, and by HEC, PK.}
\markboth{IEEE-DATA  Descriptor Article Template}{Author \textit{et al.}}

\begin{abstract}
In this work, we present an unmanned aerial vehicle (UAV) wireless dataset collected as part of the AERPAW Autonomous Aerial Data Mule (AADM) challenge, organized by the NSF Aerial Experimentation and Research Platform for Advanced Wireless (AERPAW) project. The AADM challenge was the second competition in which an autonomous UAV acted as a data mule, where the UAV downloaded data from multiple base stations (BSs) in a dynamic wireless environment. Participating teams designed flight control and decision-making algorithms for choosing which BSs to communicate with and how to plan flight trajectories to maximize data download within a mission completion time. The competition was conducted in two stages: Stage 1 involved development and experimentation using a digital twin (DT) environment, and in Stage 2, the final test run was conducted on the outdoor testbed. The total score for each team was compiled from both stages. The resulting dataset includes link quality and data download measurements, both in DT and physical environments. Along with the USRP measurements used in the contest, the dataset also includes UAV telemetry,  Keysight RF sensors position estimates, link quality measurements from LoRa receivers, and Fortem radar measurements. It supports reproducible research on autonomous UAV networking, multi-cell association and scheduling, air-to-ground propagation modeling, DT-to-real-world transfer learning, and integrated sensing and communication, which serves as a benchmark for future autonomous wireless experimentation.
\\
 {\textcolor{ieeedata}{\abstractheadfont\bfseries{IEEE SOCIETY/COUNCIL}}}     IEEE Communications Society (ComSoc), IEEE Vehicular Technology
Society (VTS), IEEE Aerospace and Electronic Systems Society (AESS)\\  
 {\textcolor{ieeedata}{\abstractheadfont\bfseries{DATA DOI/PID}}}     https://doi.org/10.5061/dryad.7d7wm3898\\ 
 {\textcolor{ieeedata}{\abstractheadfont\bfseries{DATA TYPE/LOCATION}}} Time-series; AERPAW Lake Wheeler Field Labs, Raleigh, North Carolina, USA; AERPAW Digital Twin (DT) for Lake Wheeler Field Labs
\end{abstract}

\maketitle
\section*{BACKGROUND} 
Ground-based sensing systems, such as environmental monitors, motion detectors, weather stations, and long-range (LoRa) receivers or radio frequency sensors (RFS), are often deployed in remote areas where network coverage is limited or unavailable. In these settings, data are commonly retrieved through in-person site visits, which are labor-intensive and slow. This motivates the aerial experimentation and research platform for the advanced wireless (AERPAW) \cite{AERPAW} autonomous data mule (AADM) challenge, where unmanned aerial vehicles (UAVs) can autonomously collect data from distributed ground nodes without relying on fixed communication infrastructure. Although several sensing approaches are considered, USRP-based wireless nodes are selected as the main mechanism for data collection. The goal of the AADM challenge is to evaluate how autonomous UAVs can make real-time decisions about mobility and base-station selection to maximize data collection under realistic wireless constraints.

UAVs, also known as drones, offer a promising solution to data collection challenges by serving as autonomous data mules—airborne nodes that can visit remote sensor sites and collect data. Their ability to span large geographic distances, operate autonomously, and establish temporary communication links makes them well-suited to extend connectivity to remote locations. Compared to static or terrestrial infrastructure, UAV-based data mules offer on-demand and versatile connectivity, which is particularly important for applications such as environmental monitoring, disaster recovery, and wide-scale Internet of Things (IoT) networks \cite{Francesco2024}.

However, most research work with UAV data mule systems has been conducted in simulation environments that oversimplify wireless link behavior, mobility constraints, and environmental dynamics. Such oversimplifications often fail to capture the actual interference effects, path loss variability, and channel fading caused by the motion of UAVs, altitude changes, or terrain variations. Hence, algorithms and scheduling methods fine-tuned via simulation may not be easily portable to real-world (RW) deployments, limiting their reliability and scalability to RW missions \cite{Hossen2026Aerospace}.

To bridge the gap, there is a growing need for integrated experimentation facilities that integrate virtual digital twin (DT) frameworks and physical testbeds to experiment with data collection systems enabled by UAVs. Such a platform facilitates researchers in designing, testing, and iteratively optimizing mobility and communication tactics—e.g., trajectory planning, link association, and offloading. AERPAW \cite{AERPAW} delivers this two-in-one capability via its hybrid ecosystem that consists of a DT for repeatable, controlled experiments and an outdoor testbed for RW verification.

AERPAW facilitates seamless transition from simulation to deployment for end-to-end assessment of UAV-based wireless operations from the communication, mobility, and control perspectives ~\cite{gurses2024}. Reliable and effective data mule performance requires RW datasets that reflect both wireless and mobility dynamics. These datasets should include key performance metrics (KPMs), for example, throughput, signal-to-noise ratio (SNR), and UAV trajectory, speed, and altitude collected across multiple heterogeneous ground sensors. These multi-modal data enable data-driven modeling of how UAV mobility and environmental factors affect link behavior, resulting in predictive and adaptive algorithms for future autonomous UAV data mule networks.
\subsection*{AADM Challenge}
The AADM challenge is the most recent of AERPAW's UAV student competitions, which shifts from RF source localization in the AFAR challenge ~\cite{masrur11192288}) to wireless data collection and mobility scheduling. The new challenge was designed to study how UAVs can serve as data mules—aerial nodes that download data dynamically from a number of distributed ground stations within a finite time interval.

\begin{figure}
\centerline{\includegraphics[width=3.2in]{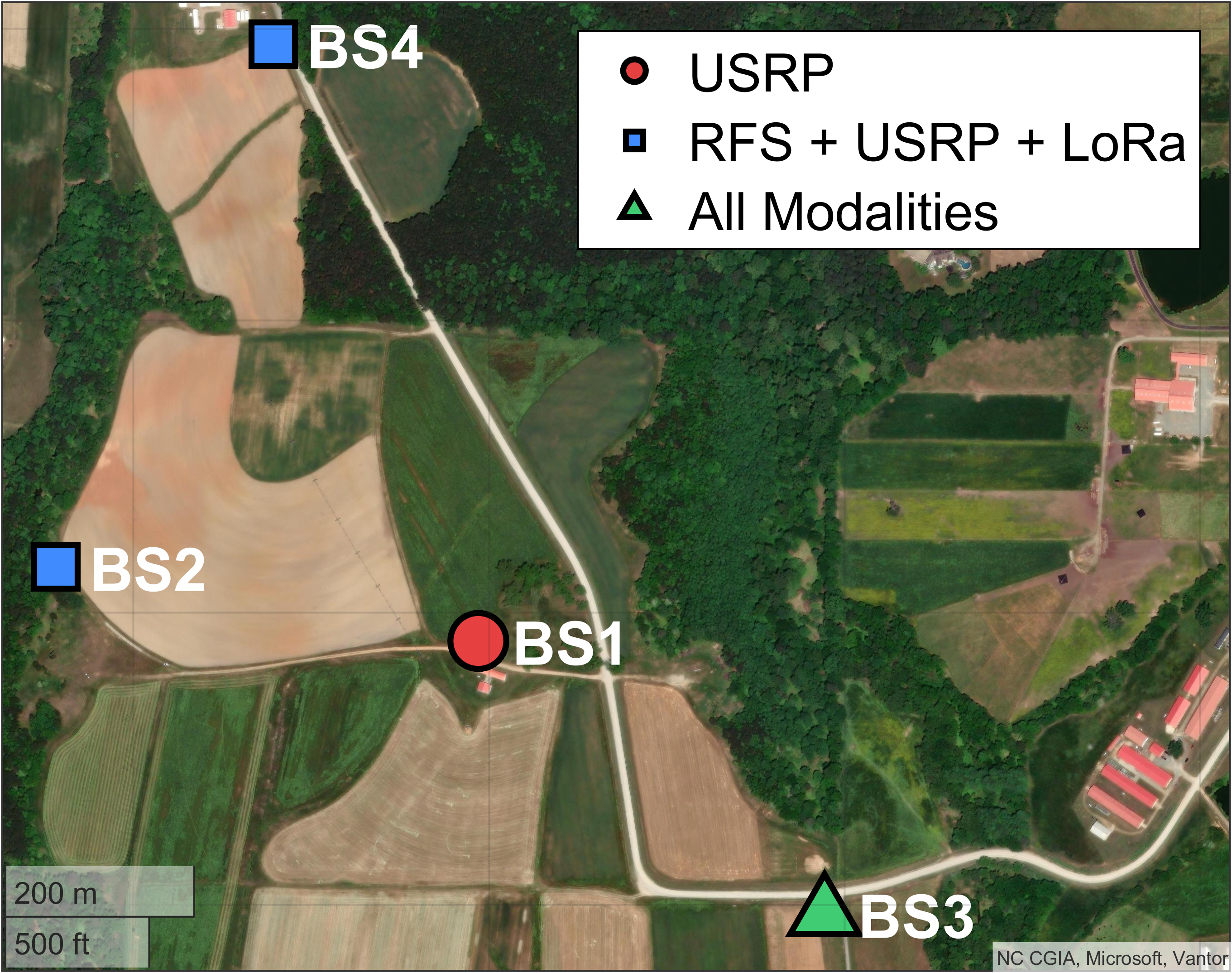}}
\caption{Experimental setup of the AADM Challenge, showing four BSs (BS1-BS4) and the sensing modalities co-located at each site across the Lake Wheeler Field Labs in Raleigh, NC. Distinct marker shapes denote the combination of sensing modalities deployed at each BS.} \label{fig:data_mule} 
\vspace{-.25in}
\end{figure}

Student teams from different universities designed UAV algorithms that could autonomously download data from multiple base stations (BSs) spread over the AERPAW Lake Wheeler Field Labs. Each UAV operated as an autonomous data mule (ADM), dynamically determining which BS to associate with, for how long, and when to switch to the next BS, to maximize total data downloaded over a 500-second flight mission. Each flight departed at an altitude of $25$ m, with four AERPAW BSs (BS-1 through BS-4) providing ground data for the Lake Wheeler Field Labs in Raleigh, as illustrated in Fig.~\ref{fig:data_mule}. Serving as an ADM, the UAV flies between BSs to collect sensor data while continuously adapting its trajectory based on link quality and mission time constraints.

The competition consisted of two stages. In Stage 1 (DT development), participants deployed and tested their algorithms on AERPAW's high-fidelity DT platform, which replicated the topology of the real testbed and wireless channel conditions. In Stage 2 (Outdoor testbed evaluation), the same containerized deployments were executed on the real testbed during field trials without any modification of the participant's code. This DT testbed architecture allowed teams to remotely develop, test, and validate ADM solutions under real communication and mobility constraints.

In addition to the USRP-based measurements used for scoring, all RW AADM flights were also monitored by external sensing instruments on the AERPAW testbed, including a network of Keysight RFS, LoRa, and a Fortem R20 radar. Although not used by competition teams, these instruments provided independent UAV tracking used for dataset enrichment. 
\subsection*{Scoring and Ranking}
Team performance was evaluated by a composite score that balanced data collection and mission time to completion. The overall score for each mission was calculated as:
\begin{equation}
    S = S_1 + S_2 - P,
    \label{eq:score}
\end{equation}
where $S_1 = (500 - T) \times I(\text{Download\_Complete})$, 
    $S_2 = 100 \times \left(\frac{D_\text{down}}{D_\text{total}}\right)$,
and $P$ represents the penalty if the UAV does not land within the specified time.  $D_\text{down}$ is the total downloaded data by the UAV from all the BSs, while $D_\text{total}$ is the total data volume set across the BSs. The final ranking was based on a weighted average of DT and testbed outcomes, with 20\% of the weight assigned to DT scores and 80\% to actual testbed results. This motivated the teams to design algorithms that generalized well from DT to RW environments, highlighting their flexibility in handling dynamic link variations and environmental uncertainty.

\subsection*{Dataset Overview and Uniqueness}
\label{sec:dataset_overview}
In an outdoor operational ADM scenario, the AADM dataset is a unique, comprehensive set of time-synchronized logs from both a DT and an RW testbed. It includes comprehensive UAV state data on heading, altitude, velocity, and position. It also provides per-second wireless data about BS association status, link quality, and throughput. In contrast to previous open datasets, it provides the first publicly available measurements of a UAV performing dynamic BS association and data-transfer scheduling based on real-time link circumstances. Moreover, AADM is the first to integrate all the multi-cell topologies in contrast to the earlier single-link AFAR \cite{masrur11192288} dataset. 
Therefore, it allows for an unprecedented opportunity to benchmark the fidelity of DTs, evaluate the transferability of algorithms across domains, and validate models for cross-layer optimization and AI-enabled resource scheduling in next-generation wireless networks.

Another distinguishing feature of the RW portion of the AADM dataset is the inclusion of three external sensing streams that were collected independently of the UAV and BSs. First, multiple Keysight RFS extracted time-difference-of-arrival (TDOA) measurements from raw I/Q data and estimated the UAV’s position as it moved across the field. In parallel, a Fortem R20 radar unit continuously tracked the UAV’s position and radial velocity while reporting radar cross-section (RCS) and additional tracking parameters throughout the missions. Finally, LoRa uplink transmissions from the UAV were received by distributed gateways across the test site, with each gateway reporting packet-level reception metadata such as received signal strength indicator (RSSI), SNR, and gateway identifiers and geolocation information. Integrating these sensing modalities with ground-truth UAV trajectory, speed, and altitude information results in a uniquely comprehensive multi-modal dataset for UAV detection and localization research.

A total of 22 teams registered for the AADM challenge, but 15 teams submitted their solutions developed in the DT to participate in the challenge. Each team designed and implemented its own UAV algorithm for ADM operations, focusing on efficient BS association and trajectory planning. 

\begin{figure}
\centerline{\includegraphics[width=3.2in,
    trim=3cm 8cm 2cm 6cm, 
    clip]{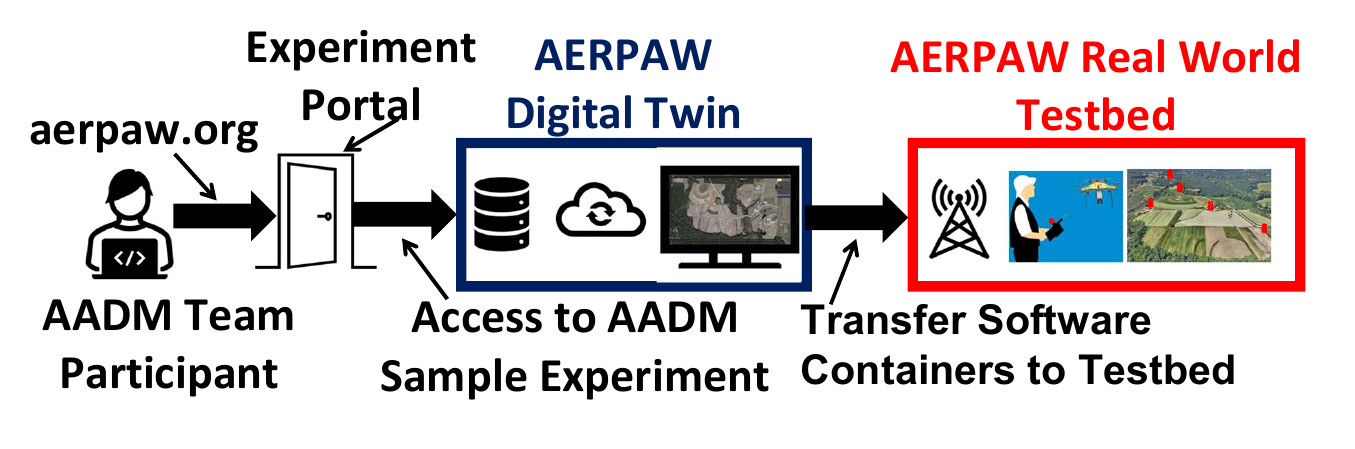}}
\caption{AADM competition workflow.\label{aadm_workflow}}
\vspace{-.3in}
\end{figure}

In the first stage (DT evaluation), each team's proposed solution was run in the DT environment with three separate UAV missions. The overall competition workflow, including the DT and RW testbed stages, is illustrated in Fig.~\ref{aadm_workflow}. Each mission corresponds to different data-volume configurations across AERPAW BSs to make the challenge more complex and to optimize the trajectory with heterogeneous traffic demands. Four data-volume configurations, denoted as D1, D2, D3, and D4, are defined across four BSs: BS1, BS2, BS3, and BS4, to introduce heterogeneous traffic demands. The first flight uses the configuration (600, 600, 600, 200) Mbits, the second uses (100, 400, 400, 100) Mbits, and the third flight uses (20, 50, 400, 30) Mbits assigned to BS1 through BS4, respectively. In each case, the UAV aims to maximize the total downloaded data within 500 seconds. In total, 45 flights (15 teams times 3 flights) are run in stage 1. 

Based on the performance in the DT, 11 finalist teams were selected for stage 2 (RW testbed evaluation).
For the RW testing, the same three-flight configuration, similar to DT, is used in the physical testbed to evaluate each team's solution, where 33 flights are run (11 teams with 3 flights each). The finalist teams were \textit{InFlux} from the University of Utah (Utah), \textit{Neural Flyers} from	North Carolina State University (NC State-1), \textit{IMPRESS@UGA} from the University of Georgia (UGA), \textit{WiMNet} from Columbia University (Columbia), \textit{NICE Lab} from North Carolina State University (NC State-2), \textit{Credente} from Florida International University (FIU), \textit{WINGS Lab} from the University at Buffalo (UB), \textit{CAAI} from  Florida Atlantic University (FAU), \textit{FlyFetchers} from National University of Sciences and Technology (NUST), \textit{
SMU ADT Lab} from Southern Methodist University (SMU), and \textit{Eagles} from the University of North Texas (UNT).
\vspace{-.1in}

\subsection*{How the Dataset can be Reused}
With an emphasis on multi-cell data collection, heterogeneous BS traffic, and autonomous association control, the AADM dataset offers a unique platform for investigating how UAV behavior is affected when communication and mobility decisions are tightly interconnected. This dataset can be reused for localization, air-to-ground (A2G) channel propagation modeling, and data-driven RW applications, as discussed in \cite{masrur11192288}. Additionally, the dataset can be further reused for the following purposes.
\vspace{-.2in}
\subsubsection*{Multi-Cell Scheduling and Association Policies}
In the AADM challenge, each experiment is evaluated in terms of three distinct BS configurations with different data volumes. This provides insights into how UAVs make dynamic scheduling and association decisions. Furthermore, this allows researchers the opportunity to explore new strategies for balancing link quality, distance, and remaining data volume to optimize total throughput and minimize flight time under RW dynamic logs. The dataset has been used for autonomous UAV trajectory design and evaluation for data muling \cite{Sadique_2025}\cite{Sadaf2026ICC} \cite{Hossen2026Aerospace}.
\vspace{-.2in}
\subsubsection*{Mission Planning Under Energy and Time Constraints}
As in the challenge, each AADM flight lasts approximately 500 seconds and captures detailed UAV motion logs, allowing for studying energy-efficient trajectory planning and time-constrained mission scheduling. Even without direct data on battery usage, researchers can explore the energy consumption from speed, acceleration, and hovering periods to investigate trade-offs between travel distance and data collection. This is important for studying how UAVs can optimize their performance in large-scale and time-sensitive data-mule missions.
\vspace{-.2in}
\subsubsection*{UAV Localization and Sensor Fusion-Based Tracking}
The inclusion of ground-truth UAV position data, position estimates from the Keysight RFS, radar tracking from the Fortem R20 unit, and LoRa gateway reception measurements enables a variety of localization and tracking studies, including ISAC-based schemes \cite{dickerson2025fusion}. Researchers can benchmark RF-based localization algorithms, evaluate tracking filters, and explore sensor fusion approaches that combine RFS, radar, and LoRa modalities. The dataset further supports analyses of bias and error characterization and the validation of theoretical bounds on localization under realistic flight geometries and channel dynamics, building on our previous work \cite{ICCDickerson}. 
\vspace{-.2in}
\subsubsection*{Fairness and Quality-of-Service Across Base Stations}
Due to the heterogeneous BS traffic with the UAVs, the dataset allows the analysis of fairness and service distribution. For example, a BS might always get neglected throughout the mission. This dataset can help explore the design of resource allocation policies similar to distributed edge computing or IoT networks.
\vspace{-.2in}
\subsubsection*{Scalability and Multi-Agent Scenarios}
It can be used to explore the scalability of data-mule operations in next-generation aerial networks. Researchers can simulate scenarios with multiple UAVs collecting data in parallel under different BS data-demand configurations and validate their results with the DT and RW testbed data logs.

\section*{COLLECTION METHODS AND DESIGN} 
The data for the AERPAW 2nd AADM Challenge was collected using AERPAW's hybrid experimentation platform, which supports both a DT and a fully programmable outdoor testbed at the Lake Wheeler Field Labs. Fig.~\ref{fig:data_mule} presents the overall system architecture used in both environments, where a UAV-mounted portable node collects the data across four fixed BSs (BS1–BS4). The physical UAV was equipped with an Intel NUC-10 computer (i7-10710U, 64 GB RAM), an onboard real-time kinematic (RTK)-enabled global navigation satellite system (GNSS) receiver, and a USRP B205 software-defined radio with a 3.4 GHz antenna for probe transmission and data-rate estimation. Each BS node was also equipped with a USRP B210 for channel sounding and executed identical receiver software. All radios were configured to operate at a 3.4 GHz center frequency with SNR updates generated each second. These hardware components provided synchronized measurements of UAV position, orientation, link quality, and the derived data-rate estimates used to compute the total amount of downloaded data. The software stack was executed in the DT, where the UAV was replaced by a software-in-the-loop (SITL) emulator and the wireless channel was modeled with a line-of-sight (LOS) plus ground-reflection propagation model, without fading or hardware imperfections. Both environments use identical message formats, experiment containers, and mission rules to ensure comparability.

The challenge provided a template for fixed and autonomous trajectory guidance. Depending on the experimenter-designed algorithm, during each flight, the UAV autonomously took off, ascended to 25 m, and received the mission-specific data volumes D1, D2, D3, and D4 from the controller. After that, data collection was conducted through a coordinated exchange between the UAV and the BS controller. The UAV's GNU radio transmitter \cite{AERPAW} periodically emitted probe signals
and the four BSs' receivers detected that probe signal and estimated individual SNR. When the UAV sent a request for link quality measurements, the controller would send the SNR value for each BS. Based on the data volume at each BS and the link quality, the UAV then starts downloading the data under the coordination of the controller. The experimenter set up their algorithm, by which the UAV could select a BS to download data in the next time interval. While the controller accumulated these data-rate estimates at each BS over time, simultaneously the UAV logged SNR, selected BS index, instantaneous rate, cumulative downloaded data, and UAV telemetry (position, velocity, attitude, and battery state). Once the 500 seconds had passed, the controller prevented the UAV from downloading data. This ensures consistency across all experiments by enforcing the download timing constraints.

After every flight, the raw logs from the UAV and the mission controller were merged and processed into the final dataset. All timestamps were converted to a common time base by using system-clock synchronization. Each mission contains time series data containing UAV GNSS position, altitude, velocity, roll/pitch/yaw angles, selected BS, instantaneous SNR for all four BSs, estimated data rate, and cumulative downloaded data. To compare DT and RW environments, log reports from both environments have the same column structure and features. The final dataset was validated with complete and consistent logs, ensuring valid geofence compliance, continuity of telemetry, and a fully executed mission.

In addition to the USRP-based communication infrastructure used for the core AADM task, Fig.~\ref{fig:data_mule} also illustrates the deployment of auxiliary sensing modalities used for independent observation of the UAV and its wireless environment. Keysight N6841A RFS co-located at BS2–BS4 passively monitored a channel-sounding waveform transmitted by the UAV at 3.32 GHz with a bandwidth of 5 MHz, serving as a proxy for control and telemetry links throughout AADM flights. A Fortem R20 radar system was mounted at a height of 10 m and co-located with BS3, operating at 16.3 GHz with a maximum bandwidth of 180 MHz to provide independent tracking measurements of the UAV. In addition, fixed LoRaWAN gateways were deployed at BS2-BS4 to enable low-power, long-range communications with a programmable USB-based LoRa end device carried by UAV.


\section*{VALIDATION AND QUALITY} 
\begin{figure}[t]
    \centering
\begin{subfigure}[t]{0.25\textwidth}
    \centering
    \includegraphics[
    width=\linewidth,
    trim=6cm 9.5cm 5.8cm 9.8cm, 
    clip
  ]{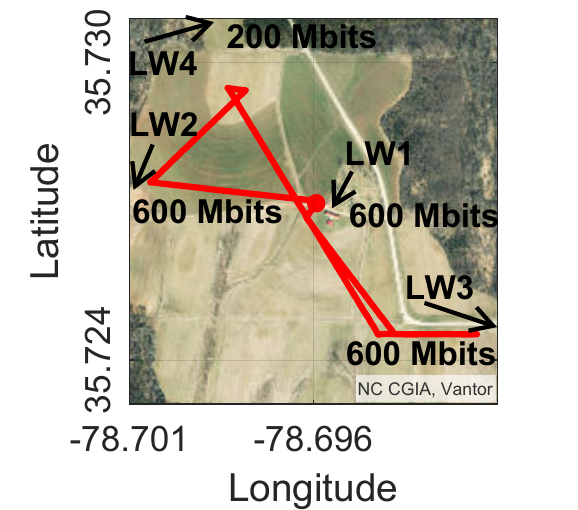}    
  \caption{Trajectory.}   
  \end{subfigure}%
  \begin{subfigure}[t]{0.23\textwidth}
    \centering
    \includegraphics[
    width=\linewidth
  ]{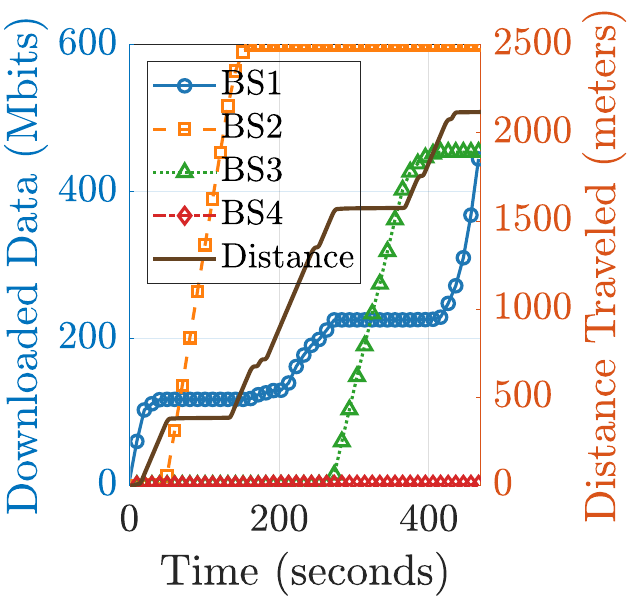}
  \caption{Download with distance vs time.}    
  \end{subfigure}%
  \caption{The controller enforced a 500-second download time limit for a given trajectory for Scenario 1, team 828.}  \label{fig:team828_traject_download}
  \vspace{-.3in}
\end{figure}

The dataset was validated to ensure that all missions in the AADM dataset contained reliable UAV telemetry, link quality measurements, and data download records. Only those flights were considered that had completed the full mission, a valid GPS fix, and intact logs from both the vehicle and the controller. For the retained missions, the UAV remained within the Lake Wheeler geofence, reached the commanded takeoff altitude of approximately 25 m, completed its traversal among the four BSs, and landed within the 500-second mission window. Following the trajectory shown in Fig.~\ref{fig:team828_traject_download}(a), the UAV downloads data from multiple BSs as illustrated in Fig.~\ref{fig:team828_traject_download}(b). We see from Fig.~\ref{fig:team828_traject_download} that the controller allows the UAV to download data only during the specified fixed download time (500 seconds). After that, it stops issuing download commands, and the cumulative downloaded data remains constant for the rest of the flight. Similarly, in the controller log, a message indicates that no further actions are required, and the system is currently off. Also, the controller outputs the final total downloaded data and the computed mission score. The cumulative download curve and the associated log messages in the controller log ensure that the 500-second download limit is strictly enforced for every mission. 
Table~\ref{tab:aadm_snr_rw} summarizes the statistics of SNR and UAV speed for a representative team, ~$799$, where the SNR values range from $-70$ dB to more than $30$ dB. These were expected for the geometry of the four BSs and variations in distance and antenna orientation during flight. BS1 and BS2 mostly reach higher average SNR values when the UAV is close to them, while BS3 and BS4 often exhibit lower average SNR due to longer ranges and the RW propagation conditions. The distribution of UAV speed is observed as expected based on mission rules, where the maximum values are close to the configured speed limit of 10 m/s, while the minimum speed is close to zero when the vehicle is hovering or taking a tight turn. 

\begin{table}[t]
\centering
\caption{Statistics of SNR for four BSs across three AADM scenarios for Team~799 in the RW testbed.}
\label{tab:aadm_snr_rw}
\begin{tabular}{|l|l|lccc|}
\hline
\textbf{Feature} & \textbf{BS} & \textbf{Stat.} & \textbf{Scenario-1} & \textbf{Scenario-2} & \textbf{Scenario-3} \\
\hline
\multirow{16}{*}{\shortstack{\textbf{SNR}\\\textbf{(dB)}}}
  & \multirow{4}{*}{\textbf{BS1}}
    & Mean & 3.65  & 0.09  & 1.52  \\
  & & Std  & 10.78 & 11.15 & 10.11 \\
  & & Max  & 26.57 & 25.87 & 27.08 \\
  & & Min  & -38.17 & -46.60 & -52.11 \\
\cline{2-6}
  & \multirow{4}{*}{\textbf{BS2}}
    & Mean & 1.65  & 1.54  & -2.85 \\
  & & Std  & 11.52 & 11.21 & 5.51  \\
  & & Max  & 32.21 & 32.29 & 8.92  \\
  & & Min  & -38.82 & -19.20 & -40.16 \\
\cline{2-6}
  & \multirow{4}{*}{\textbf{BS3}}
    & Mean & -5.05 & -8.62 & -11.46 \\
  & & Std  & 15.26 & 16.76 & 18.21 \\
  & & Max  & 20.75 & 20.74 & 20.47 \\
  & & Min  & -70.52 & -67.07 & -66.61 \\
\cline{2-6}
  & \multirow{4}{*}{\textbf{BS4}}
    & Mean & -36.33 & -33.27 & -33.39 \\
  & & Std  & 13.82 & 15.65 & 15.22 \\
  & & Max  & -4.52 & -2.65 & -1.93 \\
  & & Min  & -70.43 & -66.43 & -72.10 \\
\hline
\multirow{4}{*}{\shortstack{\textbf{Speed}\\\textbf{(m/s)}}}
  & \multirow{4}{*}{\textbf{-}}
    & Mean & 3.621  & 5.205  & 3.622  \\
  & & Std  & 4.405 & 4.632 & 4.460 \\
  & & Max  & 10.267 & 10.361 & 10.320 \\
  & & Min  & 0.004 & 0.011 &  0.006 \\
  \hline
\end{tabular}
\vspace{-.2in}
\end{table}		

\section*{RECORDS AND STORAGE} 
\begin{table*}[t]
\centering
\caption{Dataset summary across the various AADM development and experimentation phases in 2025. “All” denotes the use of all sensing modalities, including USRP, LoRa, Fortem Radar, and Keysight RFS data.}
\label{tab:dataset_summary_simple}
\setlength{\tabcolsep}{5pt}
\renewcommand{\arraystretch}{1.15}
\begin{tabular}{|l|l|c|c|c|p{9cm}|}
\hline
\textbf{Phase} & \textbf{Timeline} & \textbf{Flights} & \textbf{Altitude} & \textbf{Modalities} & \textbf{Purpose}\\
\hline
Testbed & Mar-Apr & 2 & Fixed & USRP & Guide the team to prepare their approach to how the RW environment changes.
\\
\hline
Testbed & May & 2 & Fixed & USRP & Demonstrated the challenge during the ACW  workshop. \\
\hline
Development & Jun & 45 & Varied & USRP & Teams' proposed approaches were evaluated in the DT.\\
\hline
Testbed & Jun & 8 & Varied & USRP & 
Evaluate altitude effects, channel sounding, LTE transmission behavior, and the impact of different transmit antennas at the LW1.\\
\hline
Testbed & Sep & 33 & Varied & All & Teams' proposed approaches were finally evaluated in the testbed.\\
\hline
Testbed & Sep & 3 & Varied & Radar + RFS& Improved UAV localization and tracking. \\
\hline
\end{tabular}
\vspace{-.2in}
\end{table*}

Table~\ref{tab:dataset_summary_simple}  shows an overview of the dataset across the different collection phases and how the experiments progressed from controlled digital twin (DT) studies to real-world (RW) testbed evaluations. The dataset includes early testbed flights designed to expose teams to RW channel behavior using USRP-based wireless measurements. It also contains DT flights used for algorithm development. Two testbed flights were conducted to study the effect of altitude on signal strength. Finally, the testbed evaluations with 11 finalist teams were conducted with all sensing modalities, including USRP measurements, Keysight RFS, LoRa receivers, and Fortem radar. In addition, 3 subsequent experiments were conducted using only the radar and RFS modalities, in which the UAV altitude was increased, and the transmitted signal bandwidth was expanded from 150 kHz to 5 MHz to improve localization performance \cite{ICCDickerson}.

The dataset is organized into three directories, namely USRP, LoRa, RFS, and Radar, under the main directory AADM. In the USRP folder, there are two subfolders: testbed and development. In the testbed folder (Sep), based on each team’s experiment, data volume number, and flight number, there are 33 folders.
While inside the development folder (Jun), there are also 45 folders based on experiment number and data volume.
Inside each folder, there are 5 sub-folders, namely, LW1, LW2, LW3, LW4, and UAV, representing BS1, BS2, BS3, BS4, and UAV, respectively. Inside the LW1 to LW4 folders, there are four text files, namely, power\_log.txt, quality\_log.txt, responder\_log.txt, and snr\_log.txt, which represent the received signal power, the quality of the signal, the SNR sent to the controller each second, and the SNR logs, respectively. In contrast, the LW1 folder contains an additional controller\_log.txt file, which includes all the exchanged messages between the UAV, all BSs, and the OEO console. The UAV folder contains report\_log.txt and vehicleOut.txt as  discussed in Table~\ref{tab:main_file} and the fields of the processed CSV files are discussed in Table~\ref{tab:log_fields_usrp}.

However, for the SMU ADT Lab (experiment number 1050), the experimenter modified the vehicle logging script unintentionally, which resulted in stopping the storage of the vehicle log file. Hence, for this experiment, the vehicleOut.txt file is unavailable in both environments. In the dataset, there are also 12 (2+2+8) early testbed flights with the specific name of folders as mentioned in Table~\ref{tab:dataset_summary_simple}. 

The RF Sensor and Radar directory contains 37 subfolders: 33 corresponding to testbed flights collected with all sensing modalities, three additional testbed flights conducted to improve UAV localization and tracking performance, and one folder containing post-processing scripts. Each of the 36 data subfolders includes three files: (i) a copy of the vehicleOut.txt file described above, provided again for convenience; (ii) a CSV file named AADMx.csv, where x denotes the flight number, containing all Keysight RFS localization measurements, as detailed in \cite{raouf2025wirelessdatasets}; and (iii) a JSON file containing the radar tracking measurements. The radar files include parameters such as track ID, range, azimuth, elevation, radar cross section (RCS) in dBsm, radial velocity, latitude and longitude, timestamps, as detailed in Table~\ref{tab:log_fields}. The post-processing folder includes a Jupyter notebook for converting the radar JSON files into MATLAB .mat format, along with additional MATLAB scripts/helper files for loading the data and generating example plots for the RFS and radar datasets, including Fig.~\ref{fig:localization}.

The LoRa directory contains 35 subfolders. Of these, 33 correspond to individual AADM flights, each containing a LORAlog.csv file with LoRa measurements. These files record key PHY- and packet-level parameters, including carrier frequency, signal bandwidth, spreading factor, data rate index, RSSI, SNR, packet reception timestamps, and gateway geolocation information, as summarized in Table~\ref{tab:log_fields}. An additional subfolder, AADM\_Collections\_Fig\_11, contains the data and the MATLAB code used to generate Fig.~\ref{fig:LoRaRSSI}. The directory also includes a consolidated file, All\_Logs.csv, which aggregates LoRa measurements from all flights into a single CSV. Two MATLAB scripts, extract\_packets\_vars\_usage.m and the helper function extract\_packets\_vars.m, are provided to extract relevant variables from the CSV files and load them into the MATLAB workspace. The final subfolder, separate\_vehicle\_logs, contains scripts used to organize LoRa measurements by flight. Since a single vehicleOut.txt ground-truth file was recorded across all flights, these scripts align timestamps between All\_Logs.csv and the vehicle log to partition the LoRa data into the 33 per-flight subdirectories described above.

\begin{table*}[t]
\small
\centering
\caption{Selected raw and processed files with descriptions used in the AADM challenge.}
\label{tab:main_file}
\begin{tabular}{|p{3cm}|p{14.0cm}|}
\hline
\textbf{File} & \textbf{Description} \\
\hline
\texttt{controller\_log.txt} & Initial data assignments, periodic SNR, BS selection, and control data download when to start and stop.\\
\hline
\texttt{report\_log.txt} & Controller–UAV message exchanges, including periodic SNR reporting, BS selection, and cumulative downloaded data.\\
\hline
\texttt{vehicleOut.txt} & Time-stamped UAV telemetry, including latitude, longitude, altitude, orientation, battery level, and control states.\\
\hline
\texttt{vehicleOut\_snr.csv} & All fields from $vehicleOut.txt$ and each BS SNR value from $controller\_log.txt$ after converting them to CSV.\\
\hline
\texttt{vehicleOut\_down.csv} & Fields from $vehicleOut.txt$ and each BS cumulative download from $controller\_log.txt$ after converting them to CSV.\\
\hline
\end{tabular}
\vspace{-.25in}
\end{table*}

\begin{table}[t]
\centering
\caption{Fields used in \textit{vehicleOut\_snr.csv} and \textit{vehicleOut\_down.csv}.}
\label{tab:log_fields_usrp}
\begin{tabular}{|p{2.3cm}|p{5.5cm}|}
\hline
\textbf{Field} & \textbf{Description} \\
\hline
time & Timestamp of the record (approx. 1 sec interval). \\
\hline
Altitude & UAV altitude in meters. \\
\hline
BatteryVolts & UAV battery voltage in volts. \\
\hline
GPSFix & GPS fix type/quality indicator. \\
\hline
Latitude, Longitude & UAV latitude and longitude in degrees. \\
\hline
NumberOfSatellites & Number of visible satellites. \\
\hline
Pitch, Roll, Yaw & UAV pitch, roll, and heading angles in radians. \\
\hline
VelocityX, VelocityY, VelocityZ & UAV velocity components along $X$, $Y$, and $Z$ (m/s). \\
\hline
dl\_lw$N$ (USRP) & Cumulative data download from BS$N$ (Mbits). \\
\hline
snr\_lw$N$ (USRP) & SNR for the BS$N$ link in dB. \\
\hline
\end{tabular}
\vspace{-.25in}
\end{table}
\begin{table}[t]
\centering
\caption{Key fields used in \textit{radar\_data\_N.json} and \textit{LORAlog.csv}.}
\label{tab:log_fields}
\begin{tabular}{|p{2.4cm}|p{5.5cm}|}
\hline
\textbf{Field} & \textbf{Description} \\
\hline
id (Radar) & Identification number of the track. \\
\hline
range, azimuth, elevation (Radar) & Target range (m), azimuth, and elevation angle (deg). \\
\hline
rcsDbsm (Radar) & Radar cross section (dBsm, relative to 1~m$^2$). \\
\hline
radialVelocity (Radar) & Radial velocity (m/s); positive values indicate receding targets. \\
\hline
lla (Radar) & Track latitude, longitude (deg), altitude (m). \\
\hline
gpsX (Radar) & GPS time at CPI center.\\
\hline
\makecell[l]{tx\_info\_modulation\\\_bandwidth (LoRa)}
 & LoRa signal bandwidth (Hz). \\
 \hline
tx\_info\_frequency & LoRa carrier frequency (Hz). \\
\hline
\makecell[l]{tx\_info\_modulation\\\_spreadingFactor}
 & LoRa spreading factor (SF7-SF10). \\
\hline
dr (LoRa) & LoRa data rate index. \\
\hline
rx\_info\_rssi (LoRa) & Received signal strength indicator (dBm). \\
\hline
rx\_info\_snr (LoRa) & Signal-to-noise ratio (dB). \\
\hline
rx\_info\_time (LoRa) & Packet reception time (UTC). \\
\hline
rx\_info\_N (LoRa) & N is latitude, longitude (deg), or altitude (m). \\
\hline
\end{tabular}
\vspace{-0.15in}
\end{table}

\begin{figure}[t]
    \centering
\begin{subfigure}[t]{0.16\textwidth}
    \centering
    \includegraphics[
    width=\linewidth,
    trim=6cm 9.5cm 6.5cm 10cm, 
    clip
  ]{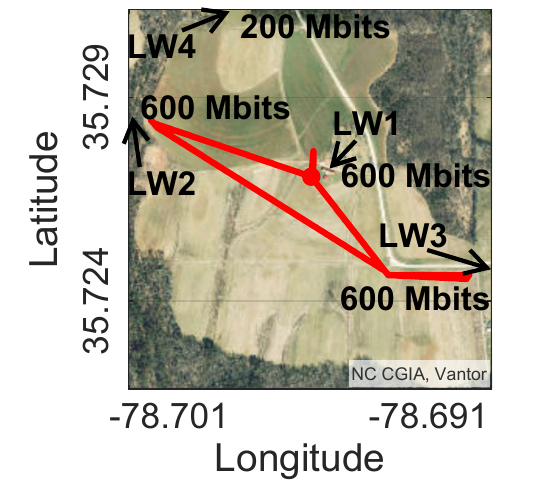}    
  \caption{Scenario 1.}    \label{fig:exp799_traj_1047_vol1}
  \end{subfigure}%
  \begin{subfigure}[t]{0.16\textwidth}
    \centering
    \includegraphics[
    width=\linewidth,
    trim=6cm 9.5cm 6.5cm 10cm, 
    clip
  ]{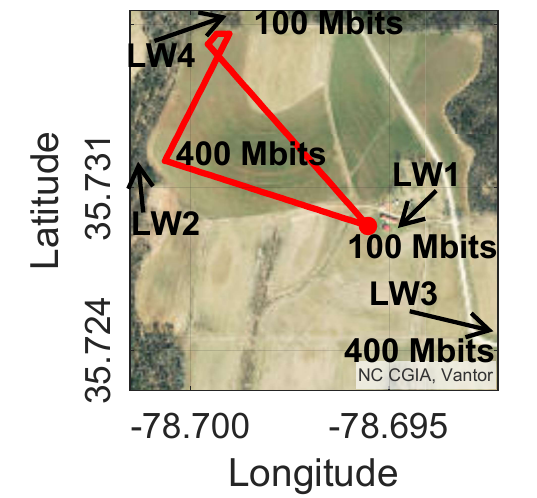}    
  \caption{Scenario 2.}
    \label{fig:exp1047_traj_vol2}
  \end{subfigure}%
  \begin{subfigure}[t]{0.16\textwidth}
    \centering
    \includegraphics[
    width=\linewidth,
    trim=6cm 9.5cm 6.5cm 10cm, 
    clip
  ]{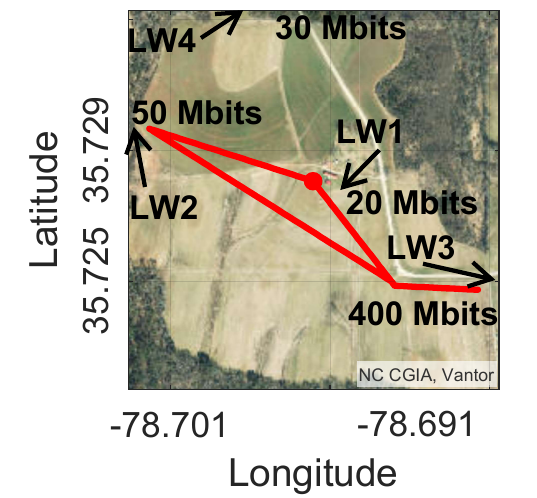}    
  \caption{Scenario 3.}    \label{fig:exp799_traj_1047_vol3}
  \end{subfigure}%
  \caption{Representative (team-1047) autonomous UAV trajectories for data download under three distinct scenarios in the development environment.}
  \label{fig:traj_devel_env}
  \vspace{-.35in}
\end{figure}  
\section*{INSIGHTS AND NOTES} 
The following representative observations are provided to guide users in interpreting the dataset and understanding common trends across UAV missions.
\begin{figure}[!h]
  \centering
  \begin{subfigure}[t]{0.22\textwidth}
    \centering    \includegraphics[width=\linewidth]{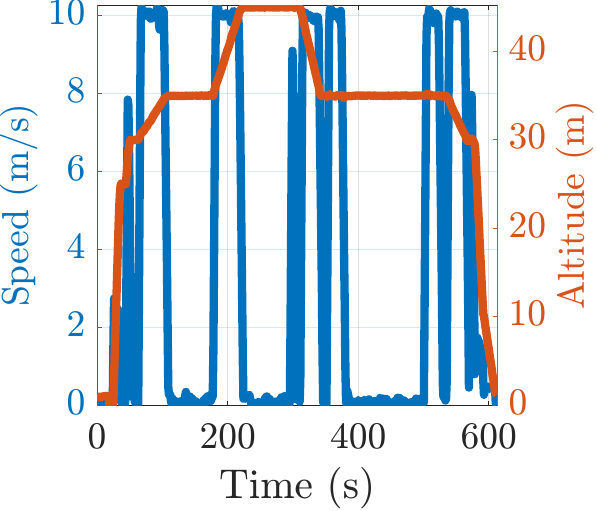}    
    \caption{Team 799}
    \label{fig:exp799_speed_alt}
  \end{subfigure}%
  \begin{subfigure}[t]{0.22\textwidth}
    \centering
    \includegraphics[width=\linewidth]{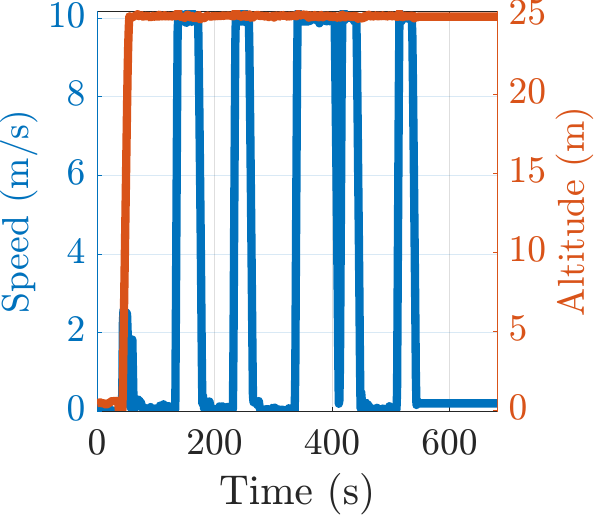}
    \caption{Team 934}
    \label{fig:exp934_speed_alt}
  \end{subfigure}%
  \hfill
  \begin{subfigure}[t]{0.22\textwidth}
    \centering
    \includegraphics[width=\linewidth]{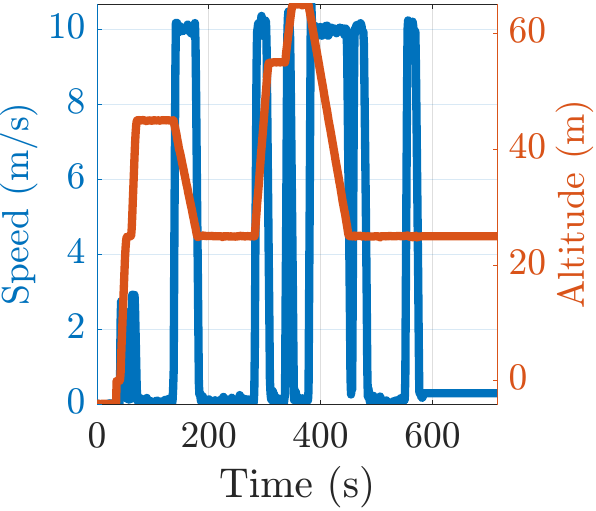}
    \caption{Team 779}
    \label{fig:exp779_speed_alt}
  \end{subfigure}
  \begin{subfigure}[t]{0.22\textwidth}
    \centering    \includegraphics[width=\linewidth]{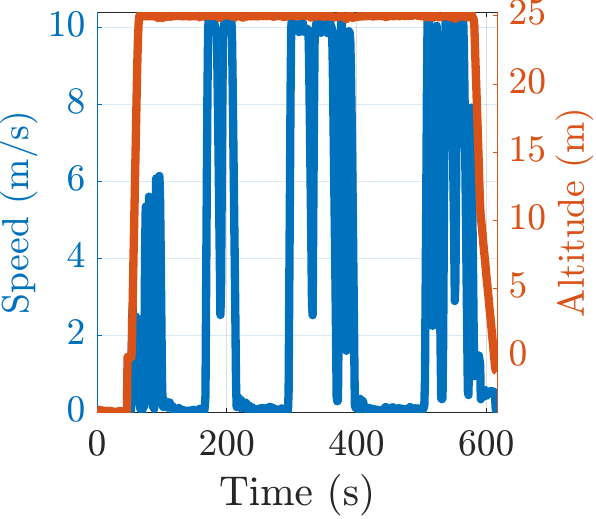}
    \caption{Team 928}
    \label{fig:exp928_speed_alt}
  \end{subfigure}
  \caption{UAV altitude and speed over time for representative teams for Scenario 1 in the outdoor environment.}
  \label{fig:speed_alt_time}
  \vspace{-.3in}
\end{figure}  

Fig.~\ref{fig:traj_devel_env} shows example autonomous UAV trajectories for representative team 1047 to download data under three different scenarios in the development environment. Based on the data volume assigned to each BS for different scenarios, the UAV follows a planned path that brings it close to multiple BSs to download data with priority. The red line indicates the flight trajectory, and the annotated values show the total data to be downloaded from each BS. It is clear from Fig.~\ref{fig:traj_devel_env} that the UAV spends more time near certain BSs to maximize the data download, while in others it follows a broader path that leads to smaller data collection.

Fig.~\ref{fig:speed_alt_time} shows the UAV altitude and speed over time for representative teams operating under the same outdoor scenario. Across all teams, the UAV ascends from its starting position at the beginning of the mission and descends near the end. Also, the UAV maintains a nearly constant altitude, with differences in the selected operating altitude reflecting distinct trajectory design choices made by each team. These altitude selections affect both LOS conditions and the effective distance to the BSs, which in turn affect the download performance.

Fig.~\ref{fig:speed_alt_time} also shows the UAV speed as a function of time, revealing a consistent pattern of acceleration and deceleration throughout the mission. The UAV accelerates while traveling toward a waypoint and decelerates upon arrival, sometimes briefly hovering before proceeding to the next waypoint. This is observed across all teams, which indicates waypoint-based navigation. Differences in the frequency and duration of speed fluctuations indicate variations in waypoint density and maneuvering strategies. Since UAV speed determines how long the UAV can associate with a BS based on the link quality, these mobility patterns play an important role in shaping data collection opportunities during the mission.

\begin{figure}[t]
  \centering
  \begin{subfigure}[t]{0.25\textwidth}
    \centering
    \includegraphics[
    width=\linewidth,
    trim=6cm 9.5cm 6cm 10cm, 
    clip
  ]{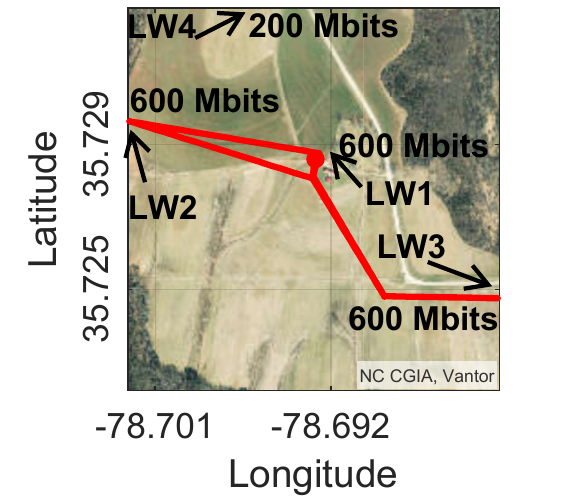}  
  \caption{Team 799}
    \label{fig:exp799_traj}
  \end{subfigure}%
  \begin{subfigure}[t]{0.25\textwidth}
    \centering
    \includegraphics[
    width=\linewidth,
    trim=6cm 9.5cm 6cm 10cm, 
    clip
  ]{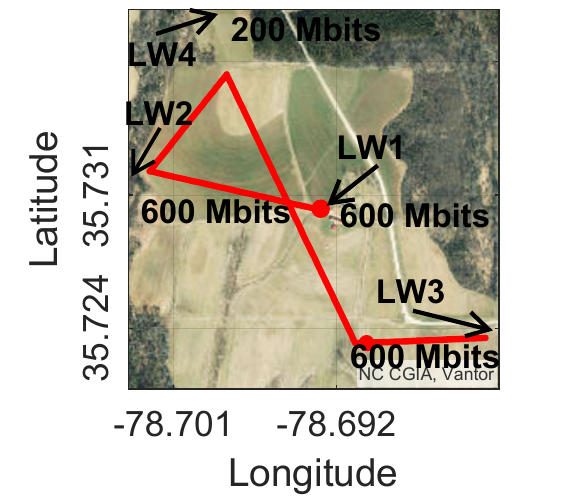}
  \caption{Team 934}
    \label{fig:exp934_traj}
  \end{subfigure}%
  \hfill  
  \begin{subfigure}[t]{0.25\textwidth}
    \centering
    \includegraphics[
    width=\linewidth,
    trim=6cm 9.5cm 6cm 10cm, 
    clip
  ]{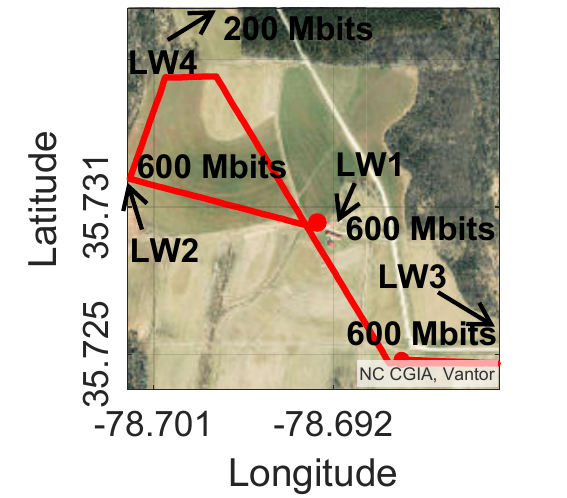}
  \caption{Team 779}
    \label{fig:exp779_traj}
  \end{subfigure}%
  \begin{subfigure}[t]{0.25\textwidth}
    \centering
    \includegraphics[
    width=\linewidth,
    trim=6cm 9.5cm 6cm 10cm, 
    clip
  ]{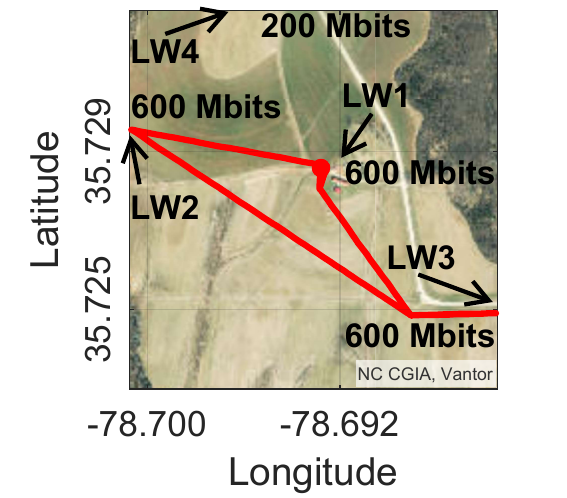}
  \caption{Team 928}
    \label{fig:exp928_traj}
  \end{subfigure}%
  \caption{UAV trajectory for representative teams for Scenario 1 in the outdoor environment.}
  \label{fig:uav_traj_outdoor}
  \vspace{-.2in}
\end{figure}
Fig.~\ref{fig:uav_traj_outdoor} shows the UAV trajectories for the representative teams under Scenario 1 in the outdoor environment. Although all teams operate within the same geographic area and mission constraints, their trajectories differ noticeably in shape and coverage. Each team adopts a distinct strategy for visiting BSs based on annotated downloaded data volumes, which is reflected in the different flight paths.
\begin{figure}[t]
  \centering  
  \vspace{-0.1in}
  \begin{subfigure}[t]{0.25\textwidth}
    \centering
    \includegraphics[
    width=\linewidth,
    trim=5.8cm 9.5cm 5cm 10cm, 
    clip
  ]{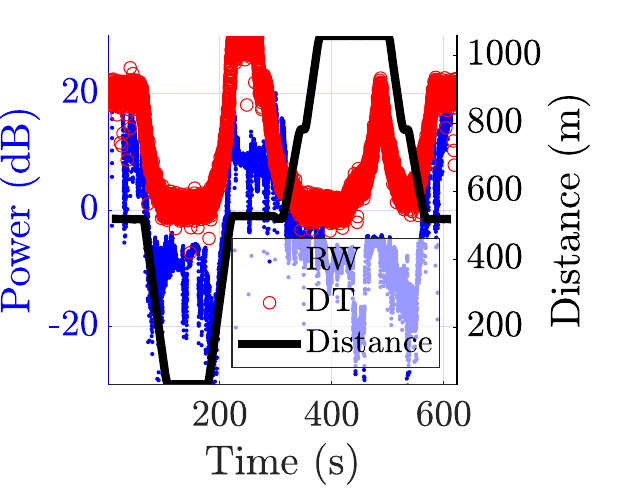}    
  \caption{Team 799}
    \label{fig:exp799_dt_rw}
  \end{subfigure}%
  \begin{subfigure}[t]{0.25\textwidth}
    \centering
    \includegraphics[
    width=\linewidth,
    trim=5.8cm 9.5cm 5cm 10cm, 
    clip
  ]{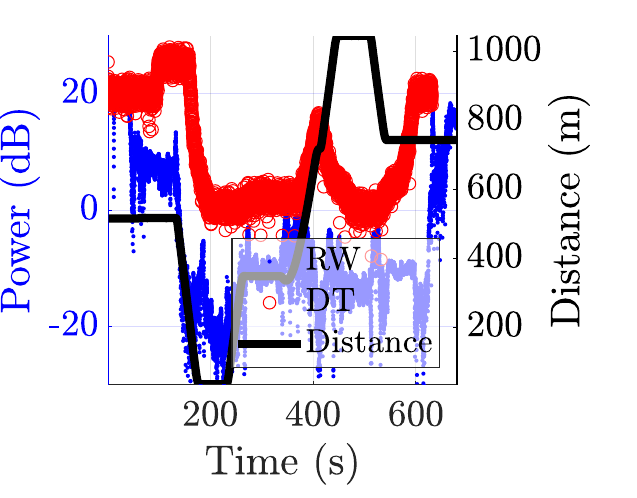}    
  \caption{Team 934}
    \label{fig:exp934_dt_rw}
  \end{subfigure}%
  \hfill
  \begin{subfigure}[t]{0.25\textwidth}
    \centering
    \includegraphics[
    width=\linewidth,
    trim=5.8cm 9.5cm 5cm 10cm, 
    clip
  ]{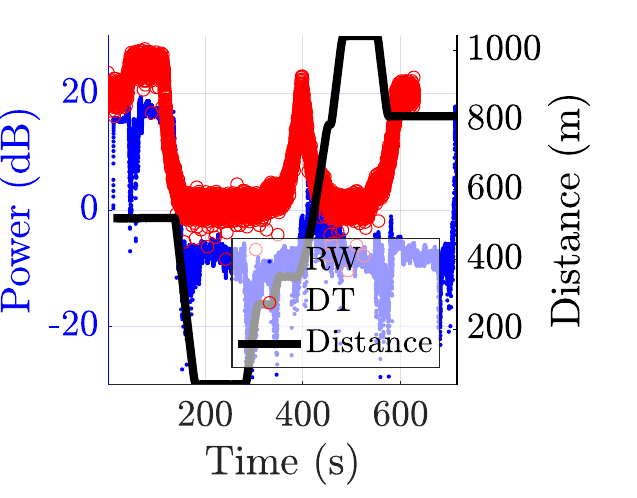}    
  \caption{Team 779}
    \label{fig:exp779_dt_rw}
  \end{subfigure}%
  \begin{subfigure}[t]{0.25\textwidth}
    \centering
    \includegraphics[
    width=\linewidth,
    trim=5.8cm 9.5cm 5cm 10cm, 
    clip
  ]{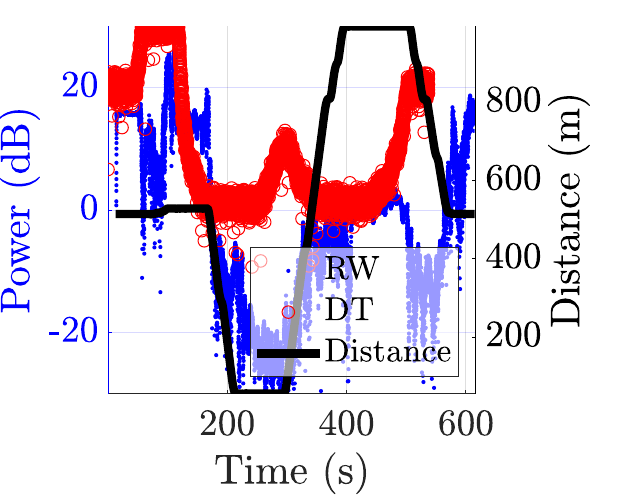}    
  \caption{Team 928}
    \label{fig:exp928_dt_rw}
  \end{subfigure}%
  \caption{Comparison of received signal power and distance over time with respect to BS1 in DT and RW for representative teams for Scenario 1 in the outdoor environment.}
  \label{fig:recv_signal_dis_time_lw1}
  \vspace{-.2in}
\end{figure}

\begin{figure}[t]
  \centering    
  \begin{subfigure}[t]{0.25\textwidth}
    \centering
    \includegraphics[
    width=\linewidth,
    trim=5.8cm 9.5cm 5cm 10.5cm, 
    clip
  ]{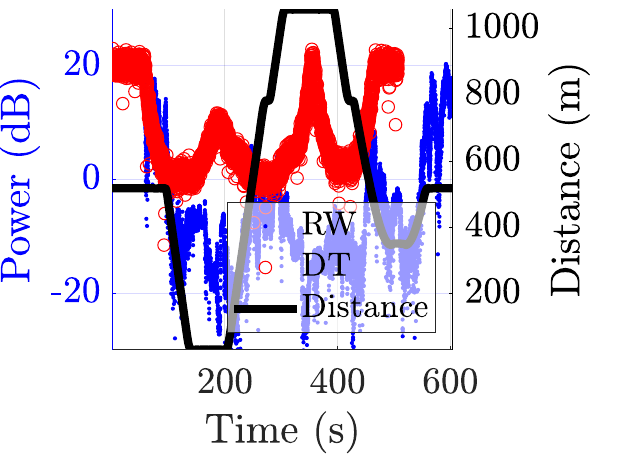}
  \caption{BS1}
    \label{fig:exp934_dt_rw}
  \end{subfigure}%
  \begin{subfigure}[t]{0.25\textwidth}
    \centering
    \includegraphics[
    width=\linewidth,
    trim=5.8cm 9.5cm 5cm 10.5cm, 
    clip
  ]{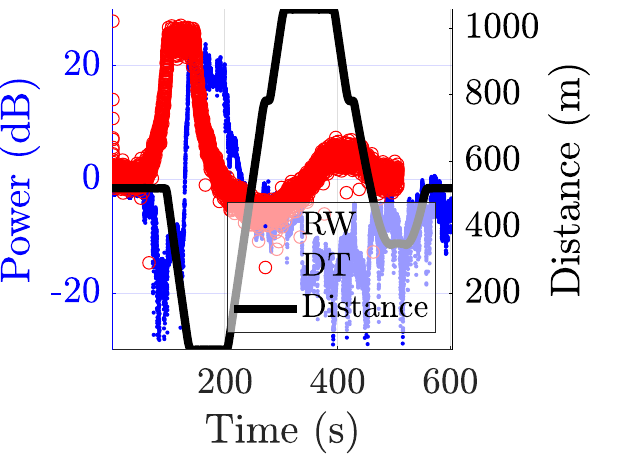}
  \caption{BS2}
    \label{fig:exp779_dt_rw}
  \end{subfigure}%
  \hfill
  \begin{subfigure}[t]{0.25\textwidth}
    \centering
    \includegraphics[
    width=\linewidth,
    trim=5.8cm 9.5cm 5cm 10.5cm, 
    clip
  ]{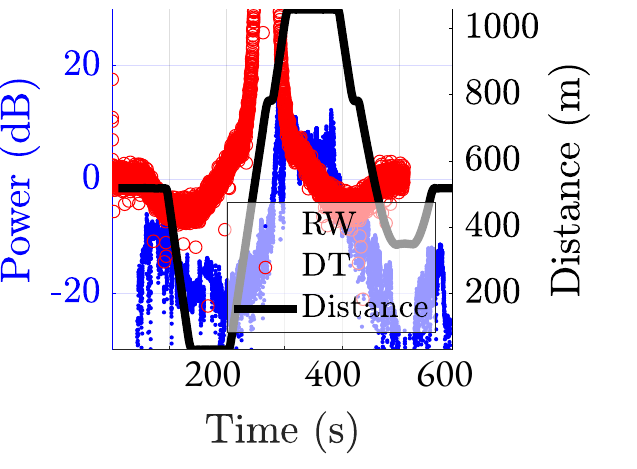}  
  \caption{BS3}
    \label{fig:exp928_dt_rw}
  \end{subfigure}%
  \begin{subfigure}[t]{0.25\textwidth}
    \centering
    \includegraphics[
    width=\linewidth,
    trim=5.8cm 9.5cm 5cm 10.5cm, 
    clip
  ]{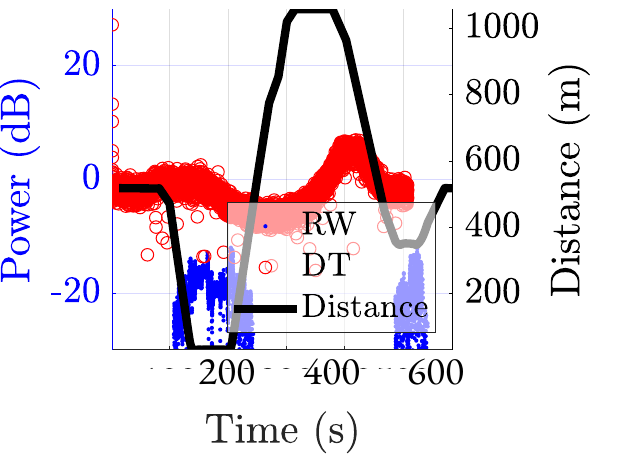}
    \caption{BS4}
    \label{fig:exp792_dt_rw}
  \end{subfigure}%
  \caption{UAV trajectory and the received signal power with respect to BS1, BS2, BS3, and BS4 for team 779 for Scenario 2 in the outdoor environment.}
  \label{fig:recv_power_dis_time_779}
  \vspace{-.3in}
\end{figure}
Fig.~\ref{fig:recv_signal_dis_time_lw1} compares the received signal power and the UAV-BS distance over time for representative teams in Scenario 1, shown for both the DT and RW environments. In all cases, received power generally decreases as the UAV moves farther from the serving BS and increases as the UAV approaches it. However, the plots also show noticeable deviations from a simple monotonic relationship. In the DT environment, received power follows a smoother trend with distance. In contrast, in the RW environment, it exhibits stronger fluctuations due to RW effects such as multipath, shadowing, antenna orientation, and environmental obstructions. These differences highlight the gap between DT and RW propagation and demonstrate the value of the dataset for studying DT-to-RW discrepancies.

Fig.~\ref{fig:recv_power_dis_time_779} shows the received signal power and UAV–BS distance over time for the representative team 779 with respect to BS1 through BS4 under Scenario 2 in the outdoor environment. For each BS, the received power increases as the UAV approaches the station and decreases as it moves away. We observe strong received power for a BS when the UAV remains nearby, but weaker power as the UAV moves away due to increased distance and geometry. However, RW measurements exhibit noticeable fluctuations compared to the smoother DT traces due to environmental effects such as multipath and shadowing.
\begin{figure}[t]
  \centering
  \begin{subfigure}[t]{0.25\textwidth}
    \centering
    \includegraphics[
    width=\linewidth,
  ]{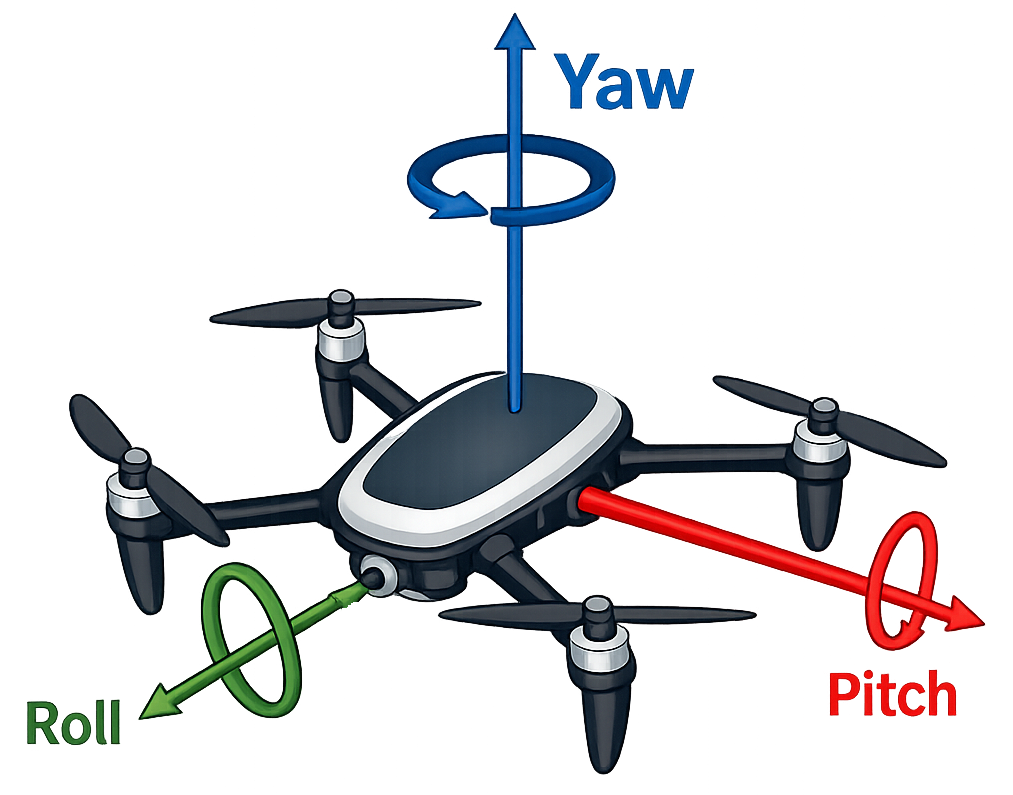}
    \caption{UAV’s roll, pitch, and yaw.}
    \label{fig:uav_rpy}
  \end{subfigure}%
  \begin{subfigure}[t]{0.2\textwidth}
    \centering
    \includegraphics[
    width=\linewidth,
  ]{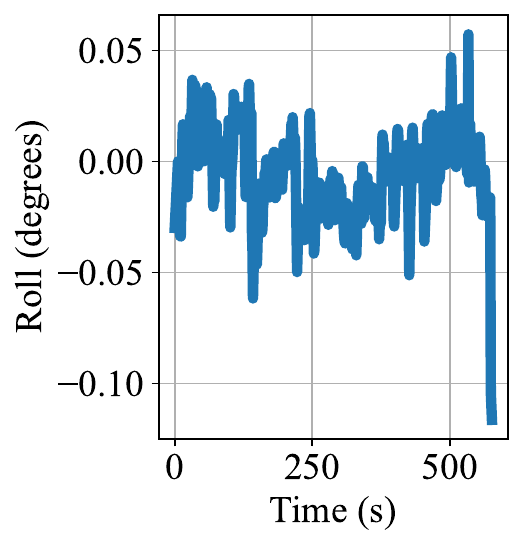}
    \caption{Roll.}
    \label{fig:uav_roll}
  \end{subfigure}%
  \hfill
  \begin{subfigure}[t]{0.2\textwidth}
    \centering
    \includegraphics[
    width=\linewidth,
  ]{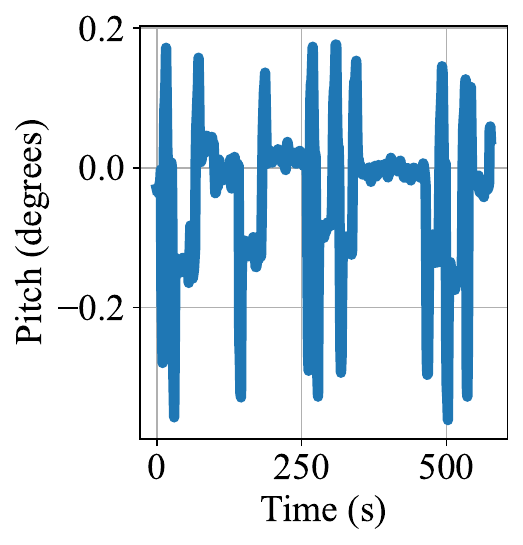}
    \caption{Pitch.}
    \label{fig:uav_pitch}
  \end{subfigure}%
  \begin{subfigure}[t]{0.2\textwidth}
    \centering
    \includegraphics[
    width=\linewidth,
  ]{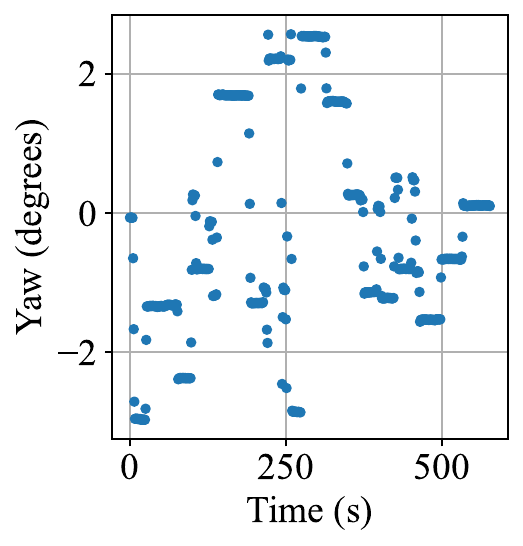}
    \caption{Yaw.}
    \label{fig:uav_yaw}
  \end{subfigure}%
  \caption{Time-series plots of the UAV’s roll, pitch, and yaw angles for Team 799 in Scenario 1.}
  \label{fig:rpy_timeseries}
  \vspace{-.3in}
\end{figure}  
Fig.~\ref{fig:rpy_timeseries} shows the UAV’s orientation, i.e., its roll, pitch, and yaw angles for Team 799 in Scenario 1 in an outdoor environment. The roll and pitch correspond to the UAV’s tilting motion, and the yaw represents changes in heading. Fig.~\ref{fig:rpy_timeseries}(b) and Fig.~\ref{fig:rpy_timeseries}(d) show that the roll and pitch remain relatively small for most of the mission, showing short spikes that may occur during turns or speed changes. However, as shown in Fig.~\ref{fig:rpy_timeseries}(d), the yaw angle changes more noticeably over time, which captures the UAV’s frequent reorientation. Although the roll and pitch angles are small in magnitude, they can still affect antenna alignment and link quality. These measurements enable the dataset to be used to study how UAV dynamics influence wireless performance in RW conditions.

 \begin{figure}
    \centering
    \includegraphics[width=0.8\linewidth]{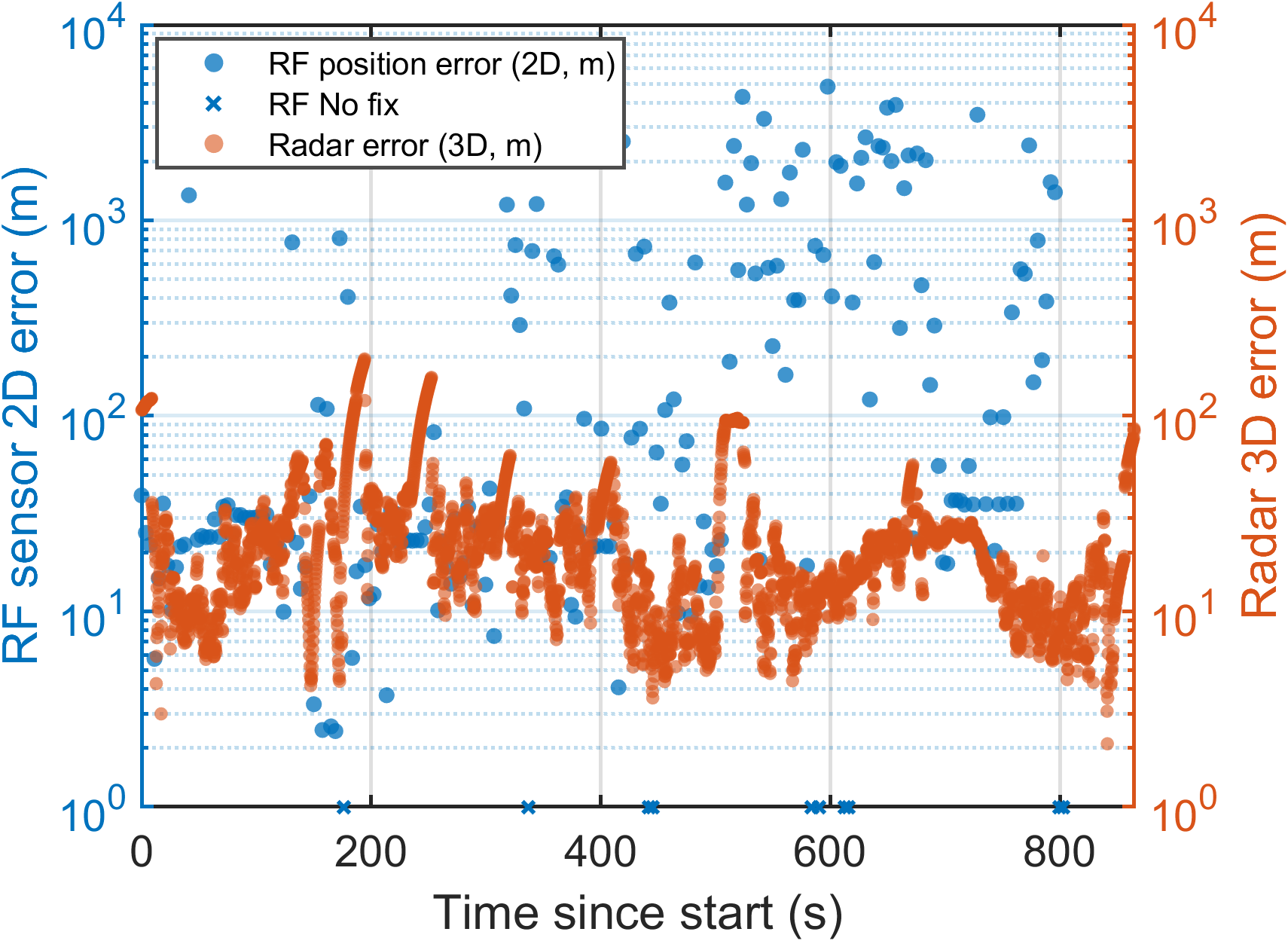}
    \caption{Localization performance comparison between the Fortem R20 radar and the Keysight N6841A RFS over time for improved localization and tracking Flight~2.}
    \label{fig:localization}
    \vspace{-.3in}
\end{figure}

\begin{figure}
    \centering
    \includegraphics[width=0.8\linewidth]{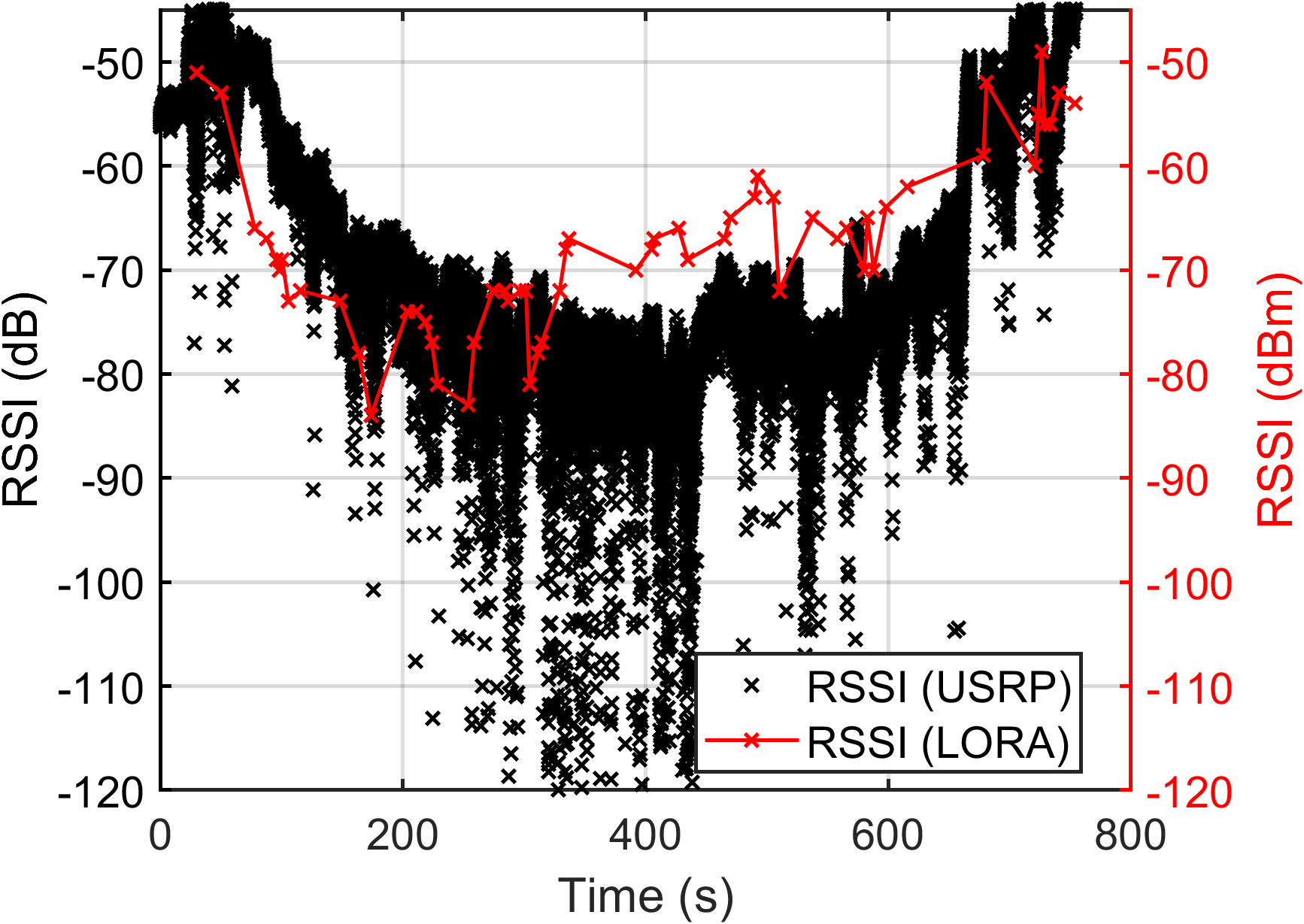}
    \caption{Normalized LoRa and USRP RSSI time series measured from AERPAW tower LW1 during a test flight conducted in September before the AADM Challenge.}
    \label{fig:LoRaRSSI}
    \vspace{-.3in}
\end{figure}

\begin{figure*}[!h]
  \centering      \includegraphics[
    width=.95\textwidth,
    trim=.1cm 10.8cm .2cm 11.2cm, 
    clip]
{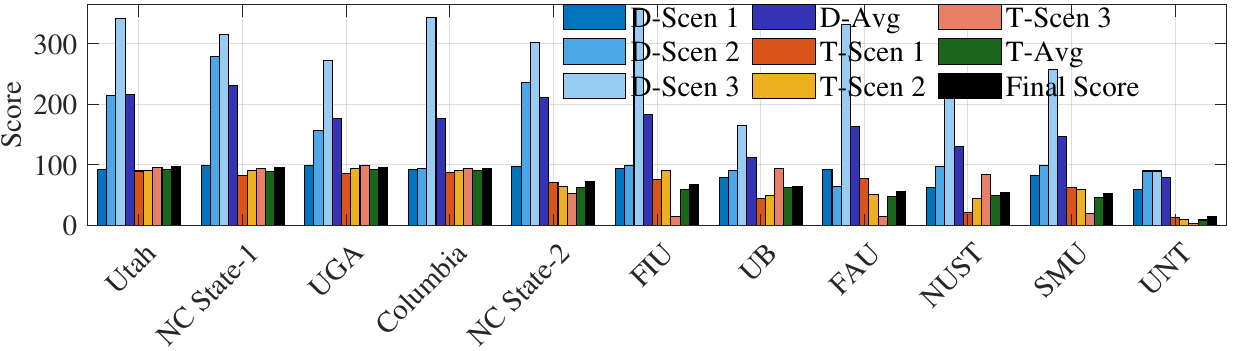}
    \caption{Scenario-wise, average and final AADM scores across teams. D and T denote DT and RW experiment modes.}
    \label{fig:aadm-scores}
    \vspace{-.3in}
\end{figure*}

 Fig.~\ref{fig:localization} illustrates RFS and radar localization over the duration of the UAV's trajectory for a representative flight. RF-based 2D position estimates exhibit variability and intermittent localization failures, but also demonstrate periods of moderate accuracy when favorable LOS and sensor geometry conditions are present, enabling wide-area, passive localization. In contrast, the radar system provides continuous 3D localization with consistently lower and more stable error. Together, these results highlight the complementary roles of RF sensing for passive, wide-area detection and radar for robust, high-accuracy tracking \cite{dickerson2025fusion}.

Fig.~\ref{fig:LoRaRSSI} shows the RSSI time series for a representative flight measured at AERPAW tower LW1, which is equipped with both a USRP and a LoRa gateway. The LoRa RSSI exhibits relatively smooth, slowly varying trends over time that are consistent with large-scale path loss and UAV trajectory geometry, in contrast to the higher-rate fluctuations and deeper fades observed in the co-located USRP RSSI. These differences are consistent with the higher carrier frequency of the USRP measurements, which are more susceptible to foliage blockage and small-scale fading, whereas the lower-frequency LoRa signals experience improved penetration through vegetation and reduced vulnerability to deep fades. Although the LoRa and USRP RSSI magnitudes are not directly comparable, and a fixed offset of -65 dB was applied to the USRP measurements for visualization purposes, the contrast in temporal behavior is robust across the trajectory. 

Finally, Fig.~\ref{fig:aadm-scores} illustrates a comparison of the AADM scores of all final participating teams. It demonstrates how each team performed across the three scenarios (denoted as Scen in Fig.~\ref{fig:aadm-scores}) in both the DT and RW testbeds under the same mission rules. We discuss all teams' details earlier in 
the \textit{Dataset Overview and Uniqueness} subsection. 
Some teams performed well across all scenarios, while others showed strong results only in specific cases and weaker performance. The figure also shows that success in a single scenario is not sufficient to achieve a high final score. A strong DT score does not guarantee a good score overall. Teams that maintained stable performance across all scenarios generally achieved higher overall rankings. This highlights the importance of designing autonomous trajectories that remain reliable across different operating conditions rather than being tuned to a single scenario.

\section*{SOURCE CODE AND SCRIPTS} All source code and post-processing scripts employed in the cleaning, merging, and analysis of the dataset are made publicly available alongside the dataset at Dryad (DOI: 10.5061/dryad.7d7wm3898). The dataset includes MATLAB source code to merge log files
to generate datasets in CSV format, compute SNR and cumulative download statistics, and recreate the trajectory, altitude, speed, and power-versus-distance plots. Other scripts are used to process the LoRa gateway logs to obtain RSSI and SNR statistics, convert the Fortem radar JSON data to a MATLAB format, and combine the Fortem, Keysight RF sensor data, and UAV ground truth data while also computing localization errors. Other LTE processing scripts are included to generate the RSRP versus distance plots from the raw IQ recordings. 

\section*{ACKNOWLEDGEMENTS AND INTERESTS}
Special thanks to NI, Keysight Technologies, AUVSI NC, and 6GNC for sponsoring the challenge awards.

This work was supported by the NSF under Grant CNS-1939334 and in part by CNS-2450593 (AERPAW). Additional support for participating teams was provided in part by NSF Grants 2332661, OAC-2512931, 2346555, CNS-2144297, CNS-2148128, CNS-2332662, CNS-2117822, EEC-2133516, CNS-2440756, SWIFT-2229563, DGE-2137100, and by HEC, PK.

M. S. Hossen prototyped the AADM challenge software framework, led the dataset curation process, processed and analyzed the USRP measurements, and prepared the initial manuscript draft. C. Dickerson processed and validated the radar and RFS datasets and assisted with manuscript preparation. O. Ozdemir processed and analyzed the LoRa and LTE measurements and validated the dataset. 

The AERPAW platform development and the execution of the field experiments were carried out by O. Ozdemir, A. Gurses, M. R. Sarbudeen, T. Zajkowski,  I. Guvenc, M. Sichitiu, M. Mushi, and R. Dutta.

All remaining authors participated in the AADM challenge and contributed to algorithm development in the digital twin environment, experimental execution, data validation, and critical review of the manuscript.

The authors declare no conflicts of interest.
\bibliographystyle{IEEEtran}
\bibliography{references}

\begin{thebibliography}{10}
\providecommand{\url}[1]{#1}
\csname url@samestyle\endcsname
\providecommand{\newblock}{\relax}
\providecommand{\bibinfo}[2]{#2}
\providecommand{\BIBentrySTDinterwordspacing}{\spaceskip=0pt\relax}
\providecommand{\BIBentryALTinterwordstretchfactor}{4}
\providecommand{\BIBentryALTinterwordspacing}{\spaceskip=\fontdimen2\font plus
\BIBentryALTinterwordstretchfactor\fontdimen3\font minus \fontdimen4\font\relax}
\providecommand{\BIBforeignlanguage}[2]{{%
\expandafter\ifx\csname l@#1\endcsname\relax
\typeout{** WARNING: IEEEtran.bst: No hyphenation pattern has been}%
\typeout{** loaded for the language `#1'. Using the pattern for}%
\typeout{** the default language instead.}%
\else
\language=\csname l@#1\endcsname
\fi
#2}}
\providecommand{\BIBdecl}{\relax}
\BIBdecl

\bibitem{AERPAW}
``{AERPAW}: Aerial experimentation and research platform for advanced wireless,'' \url{https://aerpaw.org/}, NSF Platforms for Advanced Wireless Research (PAWR), accessed: 2025-12-12.

\bibitem{Francesco2024}
F.~Betti~Sorbelli, ``{UAV}-based delivery systems: A systematic review, current trends, and research challenges,'' \emph{ACM J. Auton. Transport. Syst.}, vol.~1, no.~3, May 2024.

\bibitem{Hossen2026Aerospace}
M.~S. Hossen, O.~Ozdemir, M.~L. Sichitiu, and I.~Guvenc, ``Resilient {UAV} data mule via adaptive sensor association under timing constraints,'' in \emph{Proc. IEEE Aerosp. Conf.}, Big Sky, MT, USA, Mar. 2026.

\bibitem{gurses2024}
A.~Gurses, G.~Reddy, S.~Masrur, O.~Ozdemir, I.~Guvenc, M.~Sichitiu, A.~Sahin, A.~Alkhateeb, M.~Mushi, and R.~Dutta, ``Digital twins and testbeds for supporting {AI} research with autonomous vehicle networks,'' \emph{IEEE Comm. Mag.}, 2024.

\bibitem{masrur11192288}
S.~Masrur, O.~Ozdemir, A.~Gurses, I.~Guuvenc, M.~Sichitiu, R.~Dutta, M.~Mushi, T.~Zajkowski, C.~Dickerson, G.~Reddy, S.~V. Villar, C.-W. Wong, B.~Chatterjee, S.~Chaudhari, Z.~Li, Y.~Liu, P.~Kudyba, H.~Sun, J.~S. Mandapaka, K.~Namuduri, W.~Wang, and F.~Fund, ``Collection: {UAV}-based {RSS} measurements from the {AFAR} challenge in digital twin and real-world environments,'' \emph{IEEE Data Descr.}, pp. 1--10, 2025.

\bibitem{Sadique_2025}
J.~J. Sadique, M.~R. Khan, and A.~S. Ibrahim, ``{UAV}-aided fast data collection via machine learning using aerpaw's digital twin,'' in \emph{Proc. ACM Worksh. Wireless Netw. Testbeds}, ser. WiNTECH '25.\hskip 1em plus 0.5em minus 0.4em\relax New York, NY, USA: Assoc. Comput. Mach. (ACM), 2025, p. 89–96.

\bibitem{Sadaf2026ICC}
S.~Javed, A.~Hassan, R.~Ahmad, M.~M. Alam, M.~S. Hossen, A.~Gurses, and O.~Ozdemir, ``Autonomous uav trajectory design and evaluation for data muling in aerpaw digital twin,'' in \emph{Proc. IEEE Int. Conf. Commun. (ICC)}, Glasgow, United Kingdom, May 2026.

\bibitem{dickerson2025fusion}
\BIBentryALTinterwordspacing
C.~Dickerson, S.~Kearney, S.~Manjur, I.~Guvenc, S.~Gurbuz, A.~Gurbuz, O.~Ozdemir, and M.~Sichitiu, ``Fusion of cellular {ISAC} and passive {RF} sensing for {UAV} detection and tracking,'' 2025. [Online]. Available: \url{https://arxiv.org/abs/2512.14608}
\BIBentrySTDinterwordspacing

\bibitem{ICCDickerson}
C.~Dickerson, S.~Masrur, J.~Dickerson, O.~Ozdemir, and I.~Guvenc, ``Impact of altitude, bandwidth, and {NLOS} bias on {TDOA}-based {3D} {UAV} localization: Experimental results and {CRLB} analysis,'' in \emph{Proc. IEEE Int. Conf. Commun. Wkshps. (ICC Workshops)}, Montreal, QC, Canada, June 2025, pp. 671--677.

\bibitem{raouf2025wirelessdatasets}
\BIBentryALTinterwordspacing
A.~H.~F. Raouf \emph{et~al.}, ``Wireless datasets for aerial networks,'' 2025. [Online]. Available: \url{https://arxiv.org/abs/2510.08752}
\BIBentrySTDinterwordspacing

\end{thebibliography}
\end{document}